\documentclass[lettersize,journal]{IEEEtran}

\usepackage{cite}
\usepackage{amsmath,amssymb,amsfonts}
\usepackage{algorithmic}
\usepackage{graphicx}
\usepackage{textcomp}
\usepackage{xcolor}
\usepackage{url}
\usepackage{cleveref}
\usepackage{subcaption}
\usepackage{setspace}
\usepackage{placeins}
\usepackage{booktabs}
\usepackage{multirow}
\usepackage{gensymb}
\usepackage{commath}
\usepackage{placeins}

\usepackage{lipsum}
\usepackage{comment}

\begin{document}

\title{Predictive audio representations for early detection and tracking of hidden dynamic objects}

\author{Katerina Vinciguerra$^{1}$,~\IEEEmembership{Member,~IEEE},
        Moritz Brandes$^{2}$,
        Danilo Hollosi$^{2}$,
        and Letizia Marchegiani$^{1}$,~\IEEEmembership{Member,~IEEE}
\thanks{$^{1}$K. Vinciguerra and L. Marchegiani are with the Department of Engineering and Architecture, University of Parma, Parma, Italy
        {\tt\small \{katerina.vinciguerra, letizia.marchegiani\}@unipr.it}}%
\thanks{$^{2}$M. Brandes and D. Hollosi are with the Fraunhofer Institute for Digital Media Technology IDMT, Oldenburg, Germany.}%
}

\maketitle

\begin{abstract}
Predicting potential dangers is core to safety. Forecasting the presence of other traffic agents is core to danger prediction. Occluded traffic agents challenge detection systems as they might become \textit{visible} too late, leaving the autonomous vehicle too little time to identify, plan and act accordingly in a robust and safe way. Previous works proved that auditory perception, being omnidirectional and not constrained by a \textit{field-of-view}, provides fundamental cues for early spotting of different road users, even when hidden by other vehicles or infrastructures. Yet, those contributions deal with scenarios with only one vehicle present, and they either identify the type of vehicle or estimate its direction of arrival. In this work, we move forward, and propose a multi-task system which, simultaneously, estimates the number of vehicles present, their type, and their direction of arrival. Our methodological contribution is a two-stage pipeline: a self-supervised pre-training stage inspired by the Joint- Embedding Predictive Architecture (JEPA) applied directly to multichannel raw waveforms, followed by supervised multi-task fine-tuning with a bidirectional LSTM and three classification heads. The pre-training stage trains the encoder without labels, pushing it to predict the latent representation of a future audio segment from its past context. To train and test our framework, since no suitable dataset was publicly available, we collected an \textit{ad-hoc} one covering Non-Line-Of-Sight scenarios with multiple traffic agents simultaneously operating.  Experimental evaluation shows that our method outperforms the state of the art; it also proves that our design choice allows the model to learn robust representations, which can be transferred to an unseen driving scenario, maintaining reasonable and stable performance. 

\end{abstract}


\section{Introduction}
\label{sec:intro}

\IEEEPARstart{S}{afe} navigation requires anticipating road users that are not yet visible. 
Cameras and LiDARs can only operate when there is line of sight. Although the same does not apply to automotive radars, able to penetrate non-conductive materials like plastic bumpers and glass, they still cannot see through the metallic bodies of potential in-between vehicles, resulting in a limited \textit{field of view}. Sound propagates, and encodes spatial information, easily around obstacles, making acoustic perception extremely valuable in non-line-of-sight (NLOS) scenarios, such as vehicles approaching a T-junction from a hidden street.

Prior investigations demonstrated that a microphone array can estimate the presence of an approaching vehicle before it becomes visible \cite{schulz2021hearing, li2023sound, hao2024acoustic, jeon2025nlos}. These results are encouraging: yet, these works address scenarios where a \textit{single vehicle} is present, and while they recover the direction of arrival, they do not identify the \textit{type} of vehicle. A system operating in the real world, though, interacts with multiple simultaneous traffic agents: to do so safely and robustly, it needs to know the location and also the nature of those agents. Different types of vehicles, indeed, have different performance characteristics, kinematics, and movement limits (\textit{e.g.}, a city bus has a much wider turning radius and slower manoeuvrability than a car), which strongly condition planning and risk assessment. 

Acoustic traffic monitoring systems \cite{damiano2024can} provide those estimates, but operate at minute-level granularity on fixed roadside infrastructure, thus they are not directly \textit{usable} by an autonomous vehicle.

\begin{figure}[t]
    \centering
    \includegraphics[width=\columnwidth]{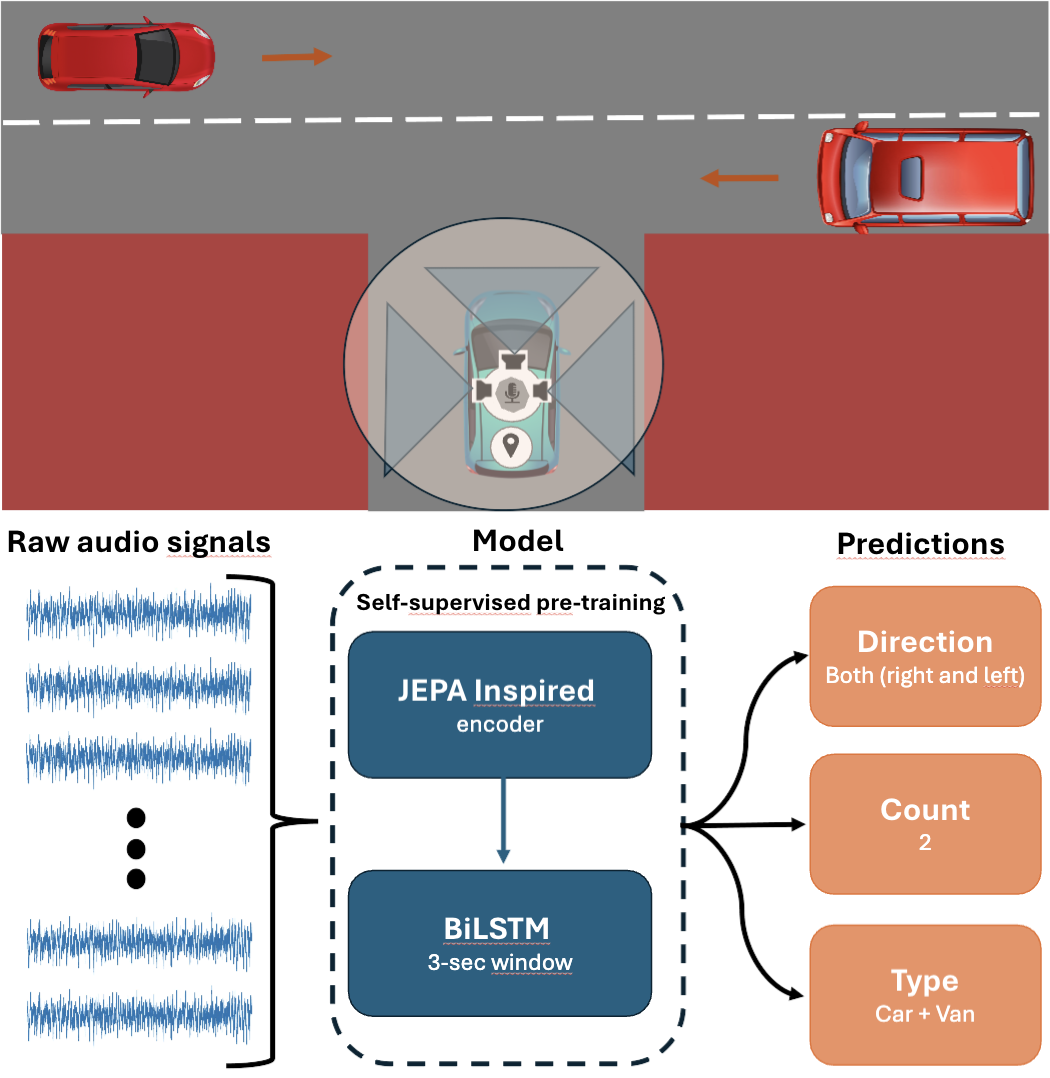}
    \caption{Example of a typical use-case: a car is approaching from the left and a van is approaching from the right at T-junction with buildings occluding them. The 8-microphone on-board array captures their raw audio signatures, which a self-supervised pretrained encoder maps into latent embeddings. A supervised head then aggregates a 3-second window of these embeddings to predict, at every second, the direction, number, and type of the non-line-of-sight vehicles.}
    \label{fig:overall_pipeline}
\end{figure}

This letter addresses this gap. We present a multi-task acoustic perception framework that simultaneously classifies, at one-second resolution, the direction of arrival (DoA) of sound sources outside the field-of-view
(\textit{e.g.}, left, right), and predicts the number of vehicles in the scene, together with their class, by using only the raw waveform from an 8-microphone circular array.  A schematic representation of the system is provided in Fig. \ref{fig:overall_pipeline}. While we acknowledge that handling the simultaneous presence of only $2$ vehicles may appear as a limitation of our approach, the challenges that arise in multi-vehicle scenes are mainly related to the ability to detect and parse the presence of \textit{more than one} simultaneous spectral signature. Our goal is to provide a proof-of-concept that it is possible to develop a multi-task pipeline able to handle planar localisation, classification and vehicle counting without relying on any sound-source separation method. Sound source separation, indeed, is pretty much still an open problem, especially in \textit{the wild} \cite{araki202530}.   Such a proof-of-concept can serve as a baseline for further approaches coping with several vehicles.

Our methodological contribution is a two-stage pipeline: a self-supervised pre-training stage inspired by the Joint-Embedding Predictive Architecture (JEPA) \cite{lecun2022jepa, assran2023ijepa}, applied directly to multichannel raw waveforms without spectral pre-processing, followed by supervised multi-task fine-tuning with a bidirectional LSTM and three classification heads. Rather than reconstructing raw audio, the encoder is trained to predict the \textit{latent representation} of a future acoustic segment from its past context, thereby adopting a temporal prediction objective that encourages semantically meaningful structure and yields representations that are temporally stable and transferable across driving conditions.



\section{Related Work}
\label{sec:relatedwork}

\textbf{Acoustic vehicle perception.}
Audio has been explored both as a complement to visual sensing and as an independent modality for early detection of traffic agents outside the field of \textit{view} \cite{fatimah2020automatic, marchegiani2022listening}.  
The first dataset explicitly designed to address NLOS scenarios was introduced in \cite{schulz2021hearing}, where the authors used SRP-PHAT \cite{hector2000ahigh} and a linear SVM to estimate the Direction-of-Arrival (DoA) of a vehicle, expressed as {\textit{left, front, right, none}. 
Subsequent work improved direction accuracy using CNN architectures  \cite{li2023sound}, jointly predicted the approach and departure directions \cite{hao2024acoustic}, and integrated acoustic cues with bird's-eye-view spatial priors \cite{jeon2025nlos}. All these works perform only DoA estimation in single-vehicle scenarios. The DCASE community, through the \textit{Acoustic-Based Traffic Monitoring} challenge and its baseline system \cite{damiano2024can}, investigated multi-vehicle localisation and classification relying on fixed roadside arrays. Yet, this setting does not fit the requirements of an autonomous vehicle's planner: the sensors are stationary, rather than on-board, and estimates are provided every minute, making them very valuable for traffic monitoring, but not to trigger immediate interventions.


\textbf{Self-supervised audio representations.}
Contrastive methods \cite{saeed2021contrastive} and masked reconstruction approaches \cite{huang2022audiomae, baevski2020wav2vec} have shown that representations learned without labels on a large audio dataset can be reused across a broad range of downstream audio tasks.
The Joint-Embedding Predictive Architecture \cite{assran2023ijepa, bardes2024vjepa} proposes a different objective: a predictor is trained to forecast the \textit{latent representation} of a masked or future region from context, avoiding the cost of reconstructing perceptually irrelevant signal detail. Audio-JEPA \cite{tuncay2025audiojepa} adapts this principle to mel-spectrogram patches and achieves competitive results on general audio benchmarks with large amounts of training data. We adapt the JEPA temporal prediction paradigm to operate directly on multichannel raw waveforms, letting the encoder free to learn from the raw signal whichever spectral, temporal, and cross-channel structure the prediction task actually rewards. 

\smallskip
Our contribution can be summarised as follows:
\begin{enumerate}
    \item To the best of our knowledge, this is the first on-board acoustic system that jointly estimates the \textit{direction}, the \textit{number}, and the \textit{class} of \textit{multiple} NLOS vehicles, with $1$-second latency. 
    \item  We introduce a JEPA-inspired temporal-prediction objective trained directly on raw 8-channel waveforms, and show that it captures better representations than supervised models, which allows generalisation to unseen, and significantly different,  scenarios.
    
\end{enumerate}

\section{Method}
\label{sec:method}

We divide the description of the proposed framework into three parts: the temporal windowing of the input signal (Section \ref{subsec:temporal}), the self-supervised pre-training of the acoustic encoder (Section \ref{subsec:firststage}), and the supervised multi-task fine-tuning stage (Section \ref{subsec:secondstage}). 
An overview of the full pipeline is shown in Fig. \ref{fig:pipeline}.

\subsection{Temporal Windowing of the Input Signal}
\label{subsec:temporal}

Let the ego-vehicle carry a circular array of $M = 8$ microphones sampled at $f_s = 48\,\text{kHz}$. The continuous multichannel data point is segmented into non-overlapping one-second audio frames. Each frame is represented as a raw waveform tensor $\mathbf{x} \in \mathbb{R}^{M \times L}$, where $L = f_s = 48K$ samples. Each frame is normalised channel-wise to zero mean and unit variance, with no spectral transform: the model operates directly on the raw waveform. Consecutive frames are grouped into sliding windows of length $T = 3$ seconds. The window centred at frame $n$ is the ordered triple
\begin{equation}
    \mathbf{W}_n
    = \bigl(\mathbf{x}_{n-2},\;\mathbf{x}_{n-1},\;\mathbf{x}_{n}\bigr),
    \label{eq:window}
\end{equation}
where subscripts denote absolute frame indices within a fixed data point scenario. Windows do not cross the scenario boundaries.

\subsection{Self-Supervised Pre-training}
\label{subsec:firststage}
\subsubsection{Architecture}

The acoustic encoder $f_\theta : \mathbb{R}^{M \times L} \to \mathbb{R}^{D}$ maps a single one-second waveform to a latent vector of dimension $D = 256$. It consists of three strided one-dimensional convolutional blocks followed by global average pooling and a linear projection:
\begin{equation}
    \mathbf{z}
    = \mathbf{W}_{\mathrm{fc}}\;
      \mathrm{GAP}\!\Bigl(
          \phi_3\!\bigl(\phi_2\!\bigl(\phi_1({\mathbf{x}})\bigr)\bigr)
      \Bigr),
    \label{eq:encoder}
\end{equation}
where ${\mathbf{x}}$ denotes the normalised input, each block $\phi_k$ is a one-dimensional strided convolution with kernel size $9$ and stride $2$, followed by batch normalisation and a ReLU activation, producing $\{64,128,256\}$ output channels for $k\in\{1,2,3\}$ respectively. $\mathrm{GAP}$ denotes global average pooling over the time dimension, and $\mathbf{W}_{\mathrm{fc}} \in \mathbb{R}^{D \times 256}$ is the linear projection matrix. The choice $D = 256$ matches the channel depth of the final convolutional block, so no information bottleneck is introduced before the latent space.

A lightweight predictor $g_\phi : \mathbb{R}^D \to \mathbb{R}^D$ approximates the mapping from a past latent representation to a future one. It is a two-layer MLP:
\begin{equation}
    g_\phi(\mathbf{z})
    = \mathrm{Linear}_{\phi_2}\!\Bigl(
        \mathrm{ReLU}\!\bigl(
            \mathrm{Linear}_{\phi_1}(\mathbf{z})
        \bigr)
      \Bigr),
    \label{eq:predictor}
\end{equation}
where $\mathrm{Linear}_{\phi_k}$ denotes a fully-connected layer with learnable parameters $\phi_k$, both mapping $\mathbb{R}^D \to \mathbb{R}^D$.

\subsubsection{Training objective}

Following the JEPA principle \cite{lecun2022jepa, assran2023ijepa}, we maintain two encoder instances: an \emph{online} encoder $f_\theta$ updated by gradient descent, and a \emph{target} encoder $f_{\bar{\theta}}$ updated by an exponential moving average (EMA) of the online weights:
\begin{equation}
    \bar{\theta}
    \;\leftarrow\;
    m\,\bar{\theta} + (1 - m)\,\theta,
    \quad m = 0.99.
    \label{eq:ema}
\end{equation}
with momentum \textit{m} following the convention established by BYOL \cite{grill2020byol} and adopted by I-JEPA \cite{assran2023ijepa}.
The target encoder receives no gradient; it provides stable prediction targets that prevent representational collapse without requiring negative pairs \cite{grill2020byol}. The target encoder is updated by EMA after every gradient step on $\theta$.

For a window $\mathbf{W}_n$, the online encoder processes the \emph{past} frame $\mathbf{x}_{n-2}$ and the predictor is trained to forecast the target encoding of the \emph{future} frame $\mathbf{x}_{n}$:
\begin{equation}
    \mathcal{L}_{\mathrm{SSL}}
    =
    \Bigl\lVert
        g_\phi\!\bigl(f_\theta(\mathbf{x}_{n-2})\bigr)
        \;-\;
        f_{\bar{\theta}}(\mathbf{x}_{n})
    \Bigr\rVert_2^2.
    \label{eq:jepa_loss}
\end{equation}
The intermediate frame $\mathbf{x}_{n-1}$ is intentionally excluded from pre-training, forcing the encoder to capture longer-range acoustic dynamics (see also Section \ref{sec:experiments}).
Gradients flow only through $f_\theta$ and $g_\phi$; the target encoder $f_{\bar{\theta}}$ is updated solely via Eq. \eqref{eq:ema}. Pre-training runs for 100 epochs on the training split using the Adam optimizer \cite{kingma2015adam} ($\eta = 10^{-3}$, $\beta_1 = 0.9$, $\beta_2 = 0.999$, batch size 32). After pre-training, the online encoder weights $\theta$ are retained and the predictor $g_\phi$ is discarded.

\subsection{Supervised Multi-Task Fine-tuning}
\label{subsec:secondstage}

\subsubsection{Model architecture}

The pre-trained encoder $f_\theta$ is composed with a bidirectional LSTM and three classification heads (Fig. \ref{fig:pipeline}). The three frames of window $\mathbf{W}_n$ are encoded independently by the shared encoder, producing the latent sequence $\bigl(\mathbf{z}_{n-2},\mathbf{z}_{n-1},\mathbf{z}_{n}\bigr) \in \mathbb{R}^{T\times D}$ with $\mathbf{z}_{n+k} = f_\theta(\mathbf{x}_{n+k})$, $k\in\{-2,-1,0\}$.
A single-layer bidirectional LSTM with per-direction hidden dimension $H = 128$ aggregates the sequence; at each step the forward and backward hidden states are concatenated as $\overleftrightarrow{\mathbf{h}}_{n+k} = [\overrightarrow{\mathbf{h}}_{n+k};\overleftarrow{\mathbf{h}}_{n+k}] \in \mathbb{R}^{2H}$, with $2H = D = 256$ so the BiLSTM output lives in the same ambient dimension as the encoder latent. 
Only the final step $\overleftrightarrow{\mathbf{h}}_{n}$ is propagated forward, through a shared module $\mathbf{h} = \mathrm{MLP}_{\psi}(\overleftrightarrow{\mathbf{h}}_{n}) \in \mathbb{R}^{128}$ consisting of a linear projection $\mathbb{R}^{2H} \to \mathbb{R}^{128}$ with a ReLU non-linearity, regularised by dropout with probability p=0.4 applied to both its input and output.
Three linear heads $\mathbf{W}_{\mathrm{dir}} \in \mathbb{R}^{5\times 128}$, $\mathbf{W}_{\mathrm{cnt}} \in \mathbb{R}^{3\times 128}$, $\mathbf{W}_{\mathrm{type}} \in \mathbb{R}^{4\times 128}$ project $\mathbf{h}$ to logits for the five direction classes (\textit{left, front, right, none, both/back}), three count classes ($0$, $1$, $2$), and four vehicle-type classes (\textit{car, van, car\,+\,van, none}).

\subsubsection{Training objective and protocol}

Each head is trained with cross-entropy; the total loss $\mathcal{L}$ is given by
\begin{equation}
    \mathcal{L} = \lambda\,\mathcal{L}_{\mathrm{CE}}^{\mathrm{dir}} + \mathcal{L}_{\mathrm{CE}}^{\mathrm{cnt}} + \mathcal{L}_{\mathrm{CE}}^{\mathrm{type}}, \quad \lambda = 5,
    \label{eq:finetune_loss}
\end{equation}
Fine-tuning proceeds in two phases. During a warm-up of $E_w = 5$ epochs, $f_\theta$ is frozen and only $\mathrm{MLP}_{\psi}$ and the heads are trained with Adam ($\eta_{\mathrm{head}} = 10^{-3}$). Afterwards, $f_\theta$ is unfrozen and jointly optimised at $\eta_{\mathrm{enc}} = 10^{-5}$, the rate is chosen empirically to adapt the SSL representation without catastrophic forgetting. Total training runs for $E = 20$ epochs at batch size 32; the checkpoint with the highest validation direction accuracy is retained.

\begin{figure*}[th!]
    \centering
    \includegraphics[width=\textwidth]{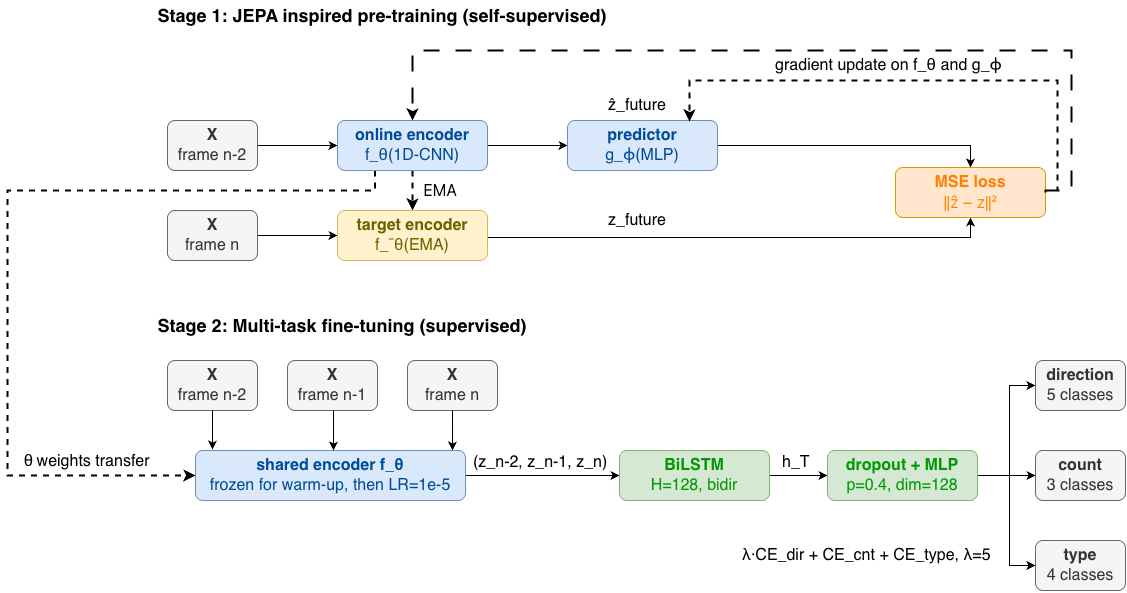}
    \caption{Overview of the proposed two-stage pipeline.
    \textbf{Stage 1 - JEPA-inspired pre-training (self-supervised):} the online encoder $f_\theta$ maps the past frame $\mathbf{x}_{n-2} \in \mathbb{R}^{M \times L}$ to a latent vector $\mathbf{z}_{n-2} \in \mathbb{R}^{D}$; the predictor $g_\phi$ forecasts the future frame $\mathbf{x}_{n}$, supervised by the target encoder $f_{\bar{\theta}}$ via $\mathcal{L}_{\mathrm{JEPA}}$ (Eq. \ref{eq:jepa_loss}); gradients flow through $f_\theta$ and $g_\phi$ only, after which $g_\phi$ is discarded and weights $\theta$ are transferred to Stage~2. 
    \textbf{Stage 2 - Multi-task fine-tuning (supervised):} the three frames of $\mathbf{W}_n$ are encoded independently, yielding $(\mathbf{z}_{n-2}, \mathbf{z}_{n-1}, \mathbf{z}_{n})$, processed by a bidirectional LSTM whose final hidden state $\overleftrightarrow{\mathbf{h}}_{n}$ passes through a shared dropout-MLP block and three linear head predicting direction (5~classes), count (3~classes), and vehicle type (4~classes) via $\mathcal{L}$ (Eq. \ref{eq:finetune_loss}, $\lambda{=}5$). }
    \label{fig:pipeline}
\end{figure*}

\section{Experiments}
\label{sec:experiments}

We run three main experiments: we evaluate the performance of our multi-task model and compare it with the state of the art (\textit{Experiment 1}); we carry out ablation studies to experimentally evaluate the rationale of our design choices (\textit{Experiment 2}); we investigate whether the model can generalise and robustly operate on unseen, and significantly different, scenarios (\textit{Experiment 3}). We indicate the three tasks: DoA estimation, vehicle type classification and number of vehicle estimation as \textit{Direction}, \textit{Count} and \textit{Type}, respectively. Some qualitative results conclude the section.

\subsection{The Dataset}
\label{subsec:dataset}
To collect real‑world data, three MER2‑503‑36U3C cameras were mounted side‑by‑side on the roof of a vehicle, directly above an octagonal array of eight microphones. A GPS unit was also installed on the roof. An overview of the setup is shown in Fig. \ref{fig:carsetting}. Each camera provides a $93$° FoV at a resolution of $2448 \times 2048$ px at $20$ fps, resulting in a combined horizontal coverage of approximately $279$°. 



\begin{figure}[!htbp]
    \centering
    \begin{minipage}[t]{0.48\linewidth}
        \centering
        \includegraphics[width=\linewidth]{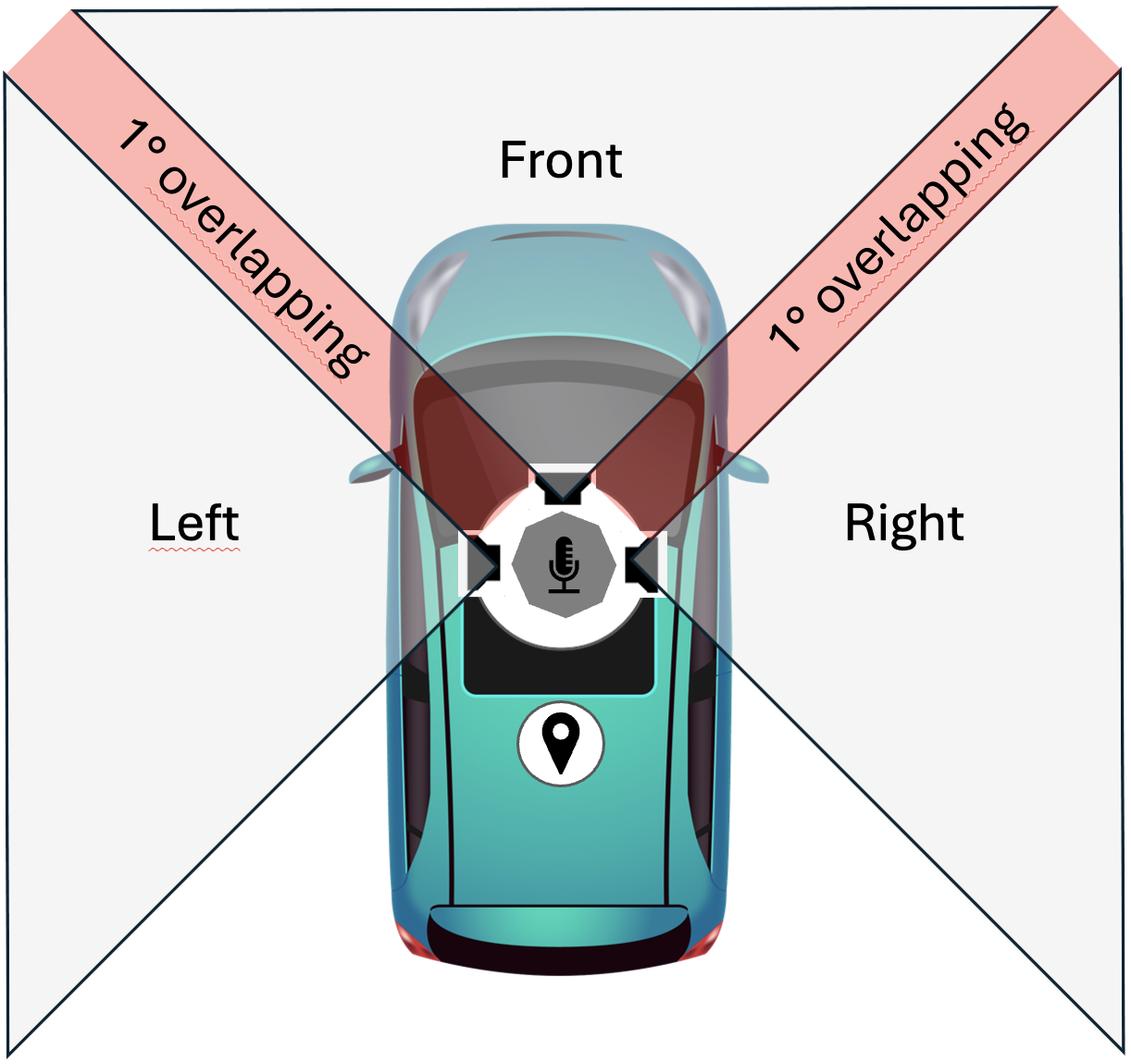}
        \label{fig:carsetting_schematic}
    \end{minipage}
    \hfill
    \begin{minipage}[t]{0.48\linewidth}
        \centering
        \includegraphics[width=\linewidth]{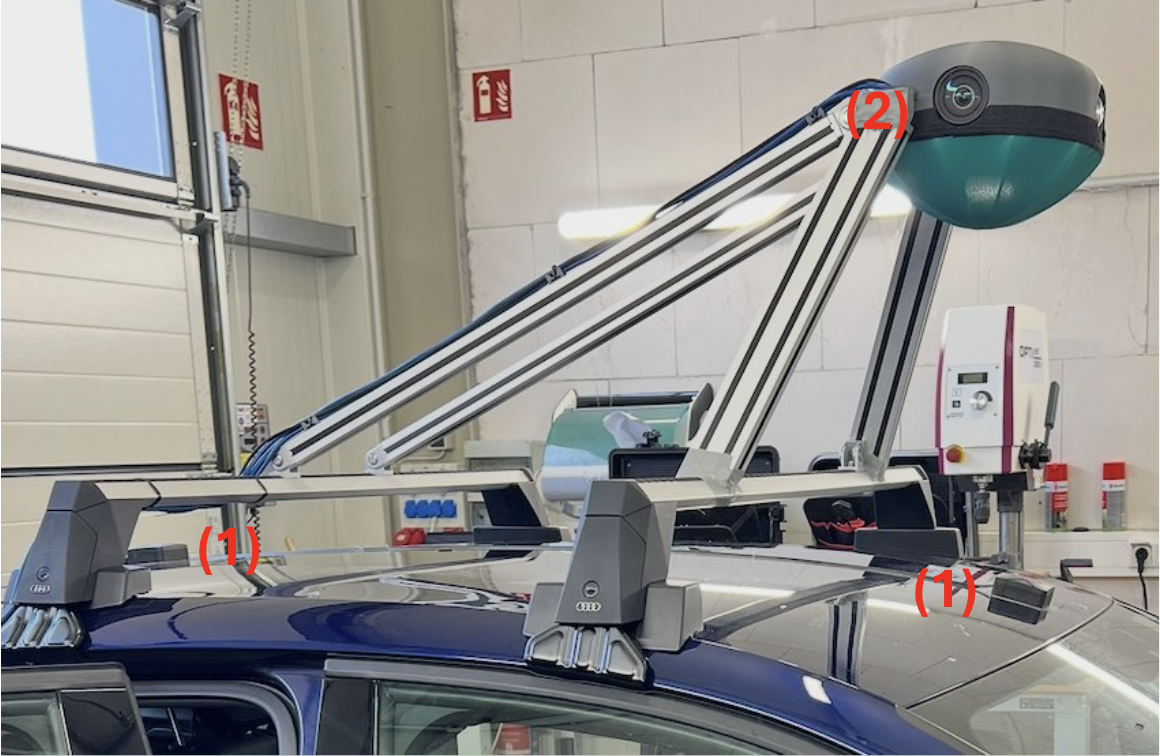}
        \label{fig:carsetting_photo}
    \end{minipage}
    \caption{Sensor configuration used in data collection. In (a): schematic top-view representation; In (b), GPS (1) and camera–microphone (2) system mounted on the vehicle. }
    \label{fig:carsetting}
\end{figure}

Data collection took place at the SolarPark in Oldenburg, Germany. The dataset spans across $6$ different scenarios (\textit{cf.}, Fig. \ref{fig:scenario}), grouped into two main settings leading to two different datasets: the one we refer to as the \textit {T-Junction (TJ)} one and the \textit{Take-Over (TO)} one. In both cases, trials last between $10$ and $30$ s, depending on the vehicle's speed and position. To reduce bias toward specific engine signatures, two different cars and two different vans were used. In the \textbf{T-junction (TJ)} dataset, the ego-vehicle is stationary at a T-junction. The dataset contains $6{,}827$ labeled audio frames of $1$s: $1{,}958$ with no vehicle present, $2{,}264$ with a single vehicle approaching (left-to-right or right-to-left), and $2{,}605$ with two vehicles approaching. In the \textbf{Take Over (TO)} dataset, 
the ego-vehicle is in motion, and one or two vehicles perform an overtake manoeuvre. The dataset contains $3{,}005$ one-second labeled audio frames: $2{,}263$ with one vehicle, and $742$ with two.

To annotate the dataset, when labels are not available directly from the experiments' design, we leveraged the cameras and the labels given by YOLOv26x \cite{sapkota2026yolo26} running on the front camera images for static scenarios and on all three camera images for the others (mistakes were manually corrected). We assign labels for all the tasks we carry on: three classes ($0$, $1$, $2$) to indicate the numer of vehicles in the scene; four classes (\textit{car, van, car+van, none}) to indicate the vehicle type, and five classes (\textit{left, front, right, none, both}) to represent the direction of arrival/departure of a vehicle. More specifically, we assign the \textit{front} label to any vehicle visible in the front camera's $93^\circ$ cone, \textit{left} to vehicles entering or leaving via the left camera, and \textit{right} via the right camera. The label \textit{both} refers to two vehicles approaching from opposite directions, the label \textit{none} indicates no vehicles. In the driving scenarios, a vehicle that has not yet reached either lateral camera is labelled as \textit{back}.

\begin{figure}[!t]
    \centering
    \setlength{\tabcolsep}{1pt}
    \begin{subfigure}[t]{0.32\linewidth}
        \centering
        \includegraphics[width=\linewidth]{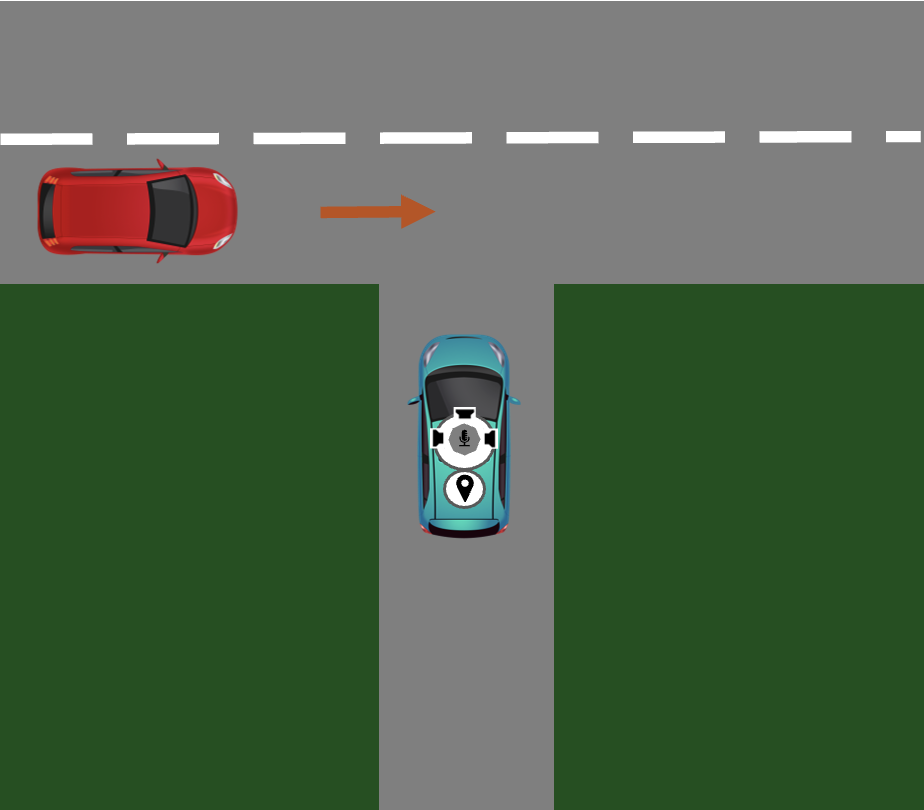}
        \caption*{(a)}
    \end{subfigure}
    \hfill
    \begin{subfigure}[t]{0.32\linewidth}
        \centering
        \includegraphics[width=\linewidth]{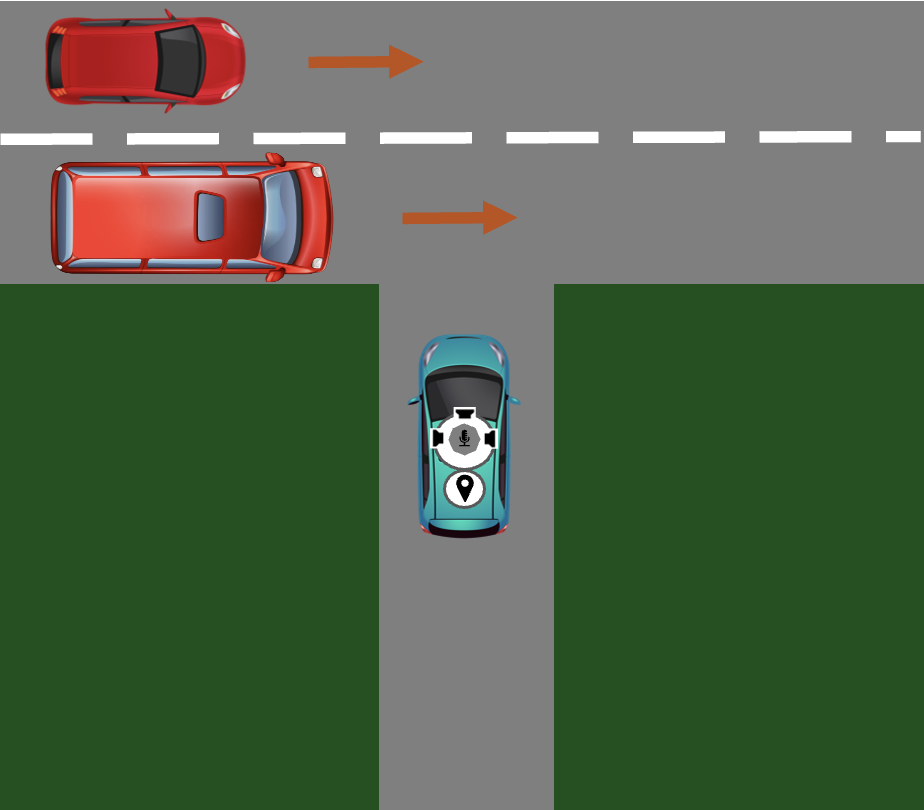}
        \caption*{(b)}
    \end{subfigure}
    \hfill
    \begin{subfigure}[t]{0.32\linewidth}
        \centering
        \includegraphics[width=\linewidth]{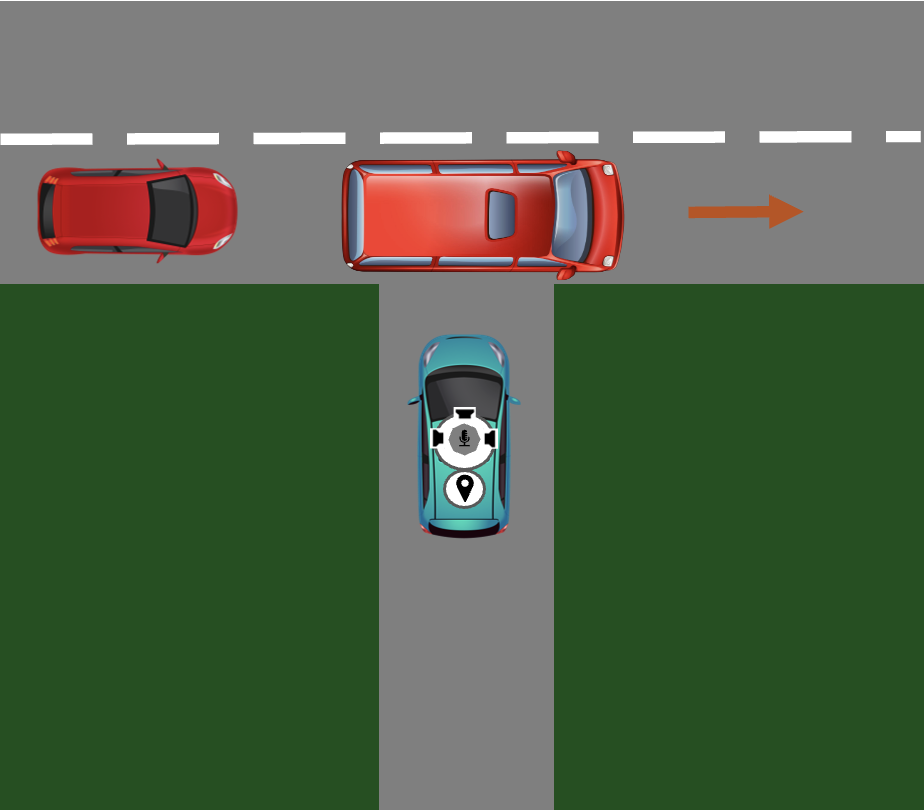}
        \caption*{(c)}
    \end{subfigure}

    \vspace{-0.1cm}

    \begin{subfigure}[t]{0.3\linewidth}
        \centering
        \includegraphics[width=\linewidth]{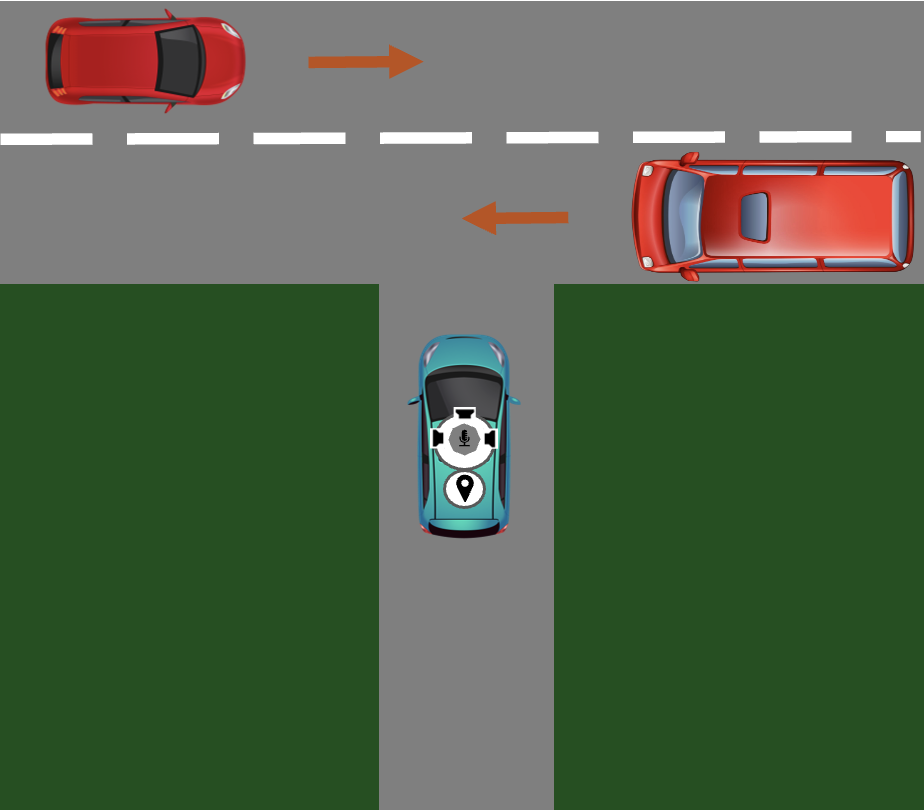}
        \caption*{(d)}
    \end{subfigure}
    \hfill
    \begin{subfigure}[t]{0.32\linewidth}
        \centering
        \includegraphics[width=\linewidth]{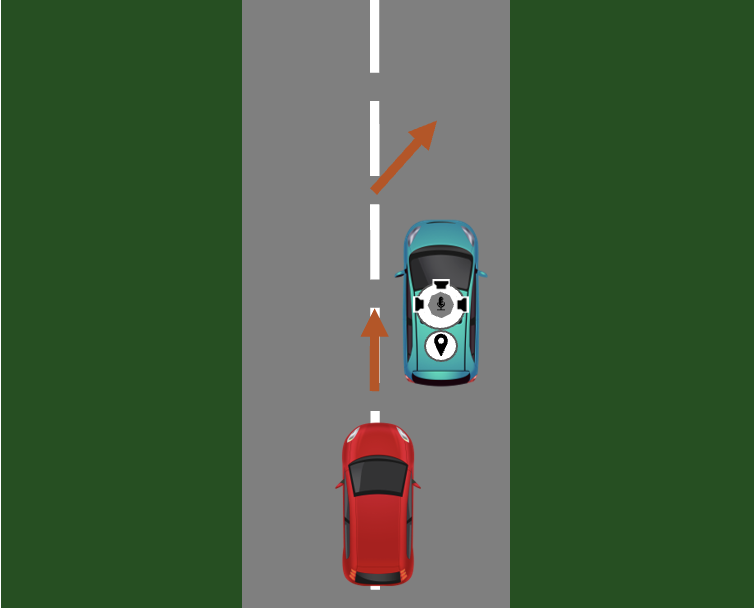}
        \caption*{(e)}
    \end{subfigure}
    \hfill
    \begin{subfigure}[t]{0.32\linewidth}
        \centering
        \includegraphics[width=\linewidth]{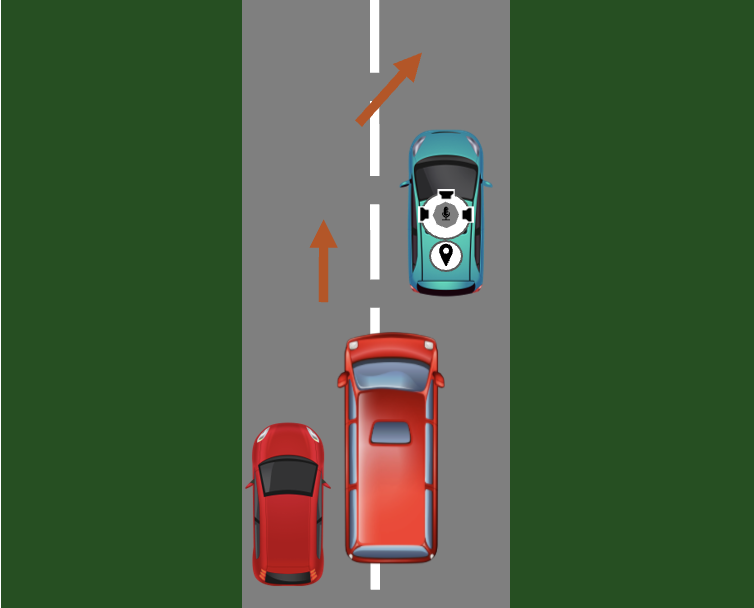}
        \caption*{(f)}
    \end{subfigure}

    \caption{Reference scenarios ; ego-vehicle in blue. \textbf{(a-d) static (T-junction):} one (a) or two vehicles (b,c) approaching from the same direction (one vehicle occluding the other); two vehicles approaching from different directions (d). \textbf{(e-f) driving (take-over):}  one (e) or two vehicles (f) approaching from the back and overtaking the ego-vehicle.}
    \label{fig:scenario}
\end{figure}

All models are trained and evaluated on the TJ dataset: $5{,}458$ frames for training, $1{,}369$ frames across $78$ for validation.  The TO dataset is held out from all training stages, including self-supervised pre-training, and is used only for the cross-scenario evaluation of Section \ref{subsec:transfer}.

\subsection{Experiment 1}
\label{subsec:multi_task}

Table \ref{tab:multitask_results} shows the performance of our model (indicated as \textit{Proposed}), and a comparison against the state of the art. We evaluate the accuracy in the \textit{Direction} tasks of the models in \cite{schulz2021hearing}, \cite{hao2024acoustic}, and \cite{li2023sound}, trained and tested on the TJ dataset (same split). Those models rely on a four-class schema for single-vehicle scenarios. Our method uses the extended five-class taxonomy, covering multiple simultaneous vehicles. Compared to the strongest prior result, even considering an additional class, our approach yields a $9.1\%$ improvement. We believe this is due to the JEPA-inspired stage of our method, capturing better representations compared to SOTA supervised models, and perform well also on smaller datasets, like ours.

As being the only other work carrying out all three tasks, we also compare ourselves against \cite{damiano2024can}, trained and tested on the TJ dataset (same split). Our model shows an improvement of $21.2\%$ in the \textit{Direction} task, $40.1\%$  in the \textit{Type} task, and $5.6\%$ in the \textit{Count} task. 
Table \ref{tab:perclass_full} (left columns) reports per-class recall on the TJ validation set. Classes \textit{none} ($94.4\%$) and \textit{both} ($100.0\%$) are the ones characterised by the highest accuracy:  the absence of any acoustic source and the simultaneous presence of two vehicles coming from both sides produce discriminating signatures. The most often confused is the \textit{front} class ($60.0\%$), with the BiLSTM reacting with some delay to the class change. For instance, a vehicle approaching from the left, whose DoA is initially, and correctly, classified as \textit{left} might be kept considered as such when it enters the \textit{front} class. More details are provided in Section \ref{subsec:front_error_analysis}. Recall in the \textit{Count} task exceeds $93\%$ across all classes, indicating that the self-supervised learning (SSL) representations encode the presence and number of distinct acoustic sources, without any architecture designed for counting. Vans achieve the highest recall ($97.0\%$) in the \textit{Type} task, likely due to their distinctive low-frequency engine signature. 

\subsection{Experiment 2}
\label{subsec:ablation}

Table \ref{tab:multitask_results} includes also the results of ablation studies run to experimentally evaluate the rationale of our design choices. We can observe that the LSTM-only baseline provides the worst performance: without the SSL inductive bias, the supervised signal alone is insufficient to learn generalised representations. We also note that the difference between frozen and unfrozen learning at stage 2 does not impact performance much, but helps better appreciate the direction task.
The \textit{Ablation - prediction-target} compares three configurations of the same encoder-predictor architecture, differing only in which frames act as context and target. The two-step horizon predicts $\mathbf{x}_n$ from $\mathbf{x}_{n-2}$; the \textit{one-step} variant predicts $\mathbf{x}_n$ from $\mathbf{x}_{n-1}$; the \textit{interpolation} variant recovers $\mathbf{x}_{n-1}$ from $\{\mathbf{x}_{n-2}, \mathbf{x}_n\}$. In the \textit{Ablation - context window}, the two-step horizon yields the best performance ($81.8\%$ direction, vs. $80.8\%$ one-step and $79.3\%$ interpolation), and is our default. One possible interpretation is that one-step prediction can be largely achieved by \textit{extending} short-time spectral content from the immediately preceding frame, providing a weaker inductive bias than the longer horizon; conversely, predicting the masked intermediate frame from a bidirectional context turns the pretext task closer to reconstruction than to forecasting.

\begin{table}[!htbp]
\centering
\caption{Results obtained in \textit{Experiment 1} and  \textit{2}, expressed as accuracy (\%) in the three tasks: \textbf{Direction}, \textbf{Count}, and \textbf{Type}. The symbol `$-$' stands for `not
applicable'.}
\label{tab:multitask_results}
\setlength{\tabcolsep}{3pt}
\begin{tabular}{p{4.0cm}ccc}
\toprule
\textbf{Method / Config.} & \textbf{Direction} & \textbf{Count} & \textbf{Type} \\
\midrule
\multicolumn{4}{l}{\textit{Baselines}} \\
\quad LSTM only             & 33.7 & 47.0 & 40.0 \\
\midrule
\multicolumn{4}{l}{\textit{Related Works}} \\
\quad DCASE baseline \cite{damiano2024can} & 67.5 & 89.5 & 66.9 \\
\quad SRP-PHAT + SVM \cite{schulz2021hearing}  & 53.4  & - & -\\
\quad DOA + TF + CNN \cite{hao2024acoustic}    & 61.9  & - & -\\
\quad Mel spec. + CNN \cite{li2023sound}       & 75.0  & - & -\\
\midrule
\multicolumn{4}{l}{\textit{Proposed}} \\
\quad SSL + BiLSTM (frozen)           & 79.9 & 94.1 & 93.2 \\
\quad SSL + BiLSTM (fine-tuned)  & \textbf{81.8} & 94.5 & \textbf{93.7} \\
\midrule
\multicolumn{4}{l}{\textit{Ablation — prediction target}} \\
\quad Interpolation & 79.3 & 89.0 & 87.3 \\ 
\quad One-step & 80.8 & 94.1 & 92.5 \\ 
\midrule
\multicolumn{4}{l}{\textit{Ablation — context window}} \\
\quad 2\,s window           & 80.4 & 93.1 & 91.1 \\
\quad 3\,s window (ours)    & \textbf{81.8} & 94.5 & \textbf{93.7} \\
\quad 4\,s window           & 81.0 & \textbf{95.3} & 93.5 \\
\bottomrule
\end{tabular}
\end{table}

\subsection{Experiment 3}
\label{subsec:transfer}

We investigate whether the model can generalise to unseen scenarios.
The TO dataset is held out from all training stages. After training on the static TJ  dataset, we compare three adaptation strategies:
\begin{itemize}
    \item \textit{Option A} (\textbf{head-only}): the SSL encoder $f_\theta$ is frozen; only the BiLSTM, shared MLP, and heads are fine-tuned.
    \item \textit{Option B} (\textbf{full-model}): all components are fine-tuned, with a small encoder learning rate ($\eta_{\mathrm{enc}}{=}10^{-5}$) to limit forgetting.
    \item \textit{Option C} (\textbf{SSL re-pretraining + head}): the temporal-prediction objective (Section \ref{subsec:firststage}) is re-applied for $30$ epochs on the unlabeled support audio, updating the encoder only; the BiLSTM and heads are then fine-tuned as in Option A.
\end{itemize}

The TO dataset introduces one new direction class, \textit{back}, absent from the static training set. The direction head is extended from $5$ to $6$ output neurons before fine-tuning, with the \textit{back} weights and bias initialised to zero; the count and type heads are unchanged. One \textit{shot} is one complete take-over data point ($10$-$30$ one-second frames). Each $(K\text{-shot}, \text{Option})$ configuration is evaluated over $N{=}10$ random draws; fine-tuning runs for $15$ epochs at $\eta_{\mathrm{head}}{=}5{\times}10^{-4}$, with the same weighting $\lambda{=}5$ as in base training. Direction recall is computed only over classes present in the TO data; \textit{none} and \textit{both} are excluded.

\subsubsection{Zero-shot transfer}
At $K{=}0$, overall direction accuracy is $2.3\%$, effectively zero by construction: the \textit{back} neuron is initialised to zero and cannot be predicted until the head receives gradient updates. Count and type accuracy collapse to $8.2\%$ and $6.5\%$, reflecting the distribution shift between stationary and dynamic conditions.

\subsubsection{Few-shot adaptation}
\label{subsec:transfer_results}
With $K{=}5$ support points, overall direction accuracy reaches $54.5$-$55.8\%$ across the three options, rising to $69.2$-$70.1\%$ at $K{=}10$ (Table \ref{tab:perclass_full}). The ranking among options is not stable across shot counts, and the spread between them is small relative to the standard deviation across the $N{=}10$ random data points. We read this as evidence that the three strategies are practically equivalent at the support sizes tested: the frozen SSL representations already suffice for driving-scenario adaptation, and neither full-model fine-tuning (Option B) nor SSL re-pretraining (Option C) measurably beats head-only adaptation (Option A). The same near-equivalence holds for count and type.

\subsubsection{Per-class behaviour under class imbalance}
Per-class recall reveals two effects of class imbalance in the TO dataset: the \textit{right} class stays persistently low ($\leq 13\%$ at both \textit{5-} and \textit{10-shot}), and the \textit{left} class is non-monotonic, declining from $32.4\%$ at \textit{5-shot} to $15.6\%$ at \textit{10-shot} under Option A. This points to two future directions: collecting more diverse data to decrease the representation gap, and developing methods for better zero-shot \textit{understanding} of long-tailed scenarios, since real-world traffic is inherently asymmetric and balanced data sets cannot be assumed.

\begin{table*}[!htbp]
\centering
\caption{Per-class recall (\%) of the proposed model on the TJ (static) validation set and on the TO (driving) dataset. Static results correspond to the best checkpoint trained on T-junction data only. Driving results are reported for the three adaptation strategies: \textit{A} (head-only), \textit{B} (full-model), \textit{C} (SSL re-pretraining + head). For $K{>}0$, each cell reports mean, $\pm$ std over $N{=}10$ random support-set draws. At zero-shot ($K{=}0$) no fine-tuning is performed and the three options are identical. \textbf{N/A} stands for `Not Available', as the class is not present in this scenario.}
\label{tab:perclass_full}
\setlength{\tabcolsep}{3pt}
\renewcommand{\arraystretch}{1.05}
\footnotesize
\begin{tabular}{llcclcccccccc}
\toprule
& & 
& \multicolumn{10}{c}{\textbf{TO Dataset}} \\
\cmidrule(lr){5-13}
& & & & \textbf{0-shot} &
  & \multicolumn{3}{c}{\textbf{5-shot}} &
  & \multicolumn{3}{c}{\textbf{10-shot}} \\
\cmidrule(lr){5-5} \cmidrule(lr){7-9} \cmidrule(lr){11-13}
\textbf{Task} & \textbf{Class} & \textbf{TJ Dataset} & & all & & \textit{A} & \textit{B} & \textit{C} & & \textit{A} & \textit{B} & \textit{C} \\
\midrule
\multirow{7}{*}{\rotatebox[origin=c]{90}{Direction}}
  & left    & 81.8  & & 1.7   & & $32.4{\pm}22.0$ & $31.4{\pm}20.9$ & $28.4{\pm}18.8$
                                & & $15.6{\pm}17.3$ & $17.7{\pm}17.6$ & $20.9{\pm}15.5$ \\
  & front   & 60.0  & & 5.6   & & $60.4{\pm}18.0$ & $60.3{\pm}17.2$ & $57.9{\pm}20.3$
                                & & $70.8{\pm}8.2$  & $70.7{\pm}8.0$  & $71.7{\pm}6.0$  \\
  & right   & 71.3  & & 13.0  & & $6.0{\pm}2.4$   & $4.3{\pm}3.2$   & $0.0{\pm}0.0$
                                & & $8.0{\pm}7.6$   & $9.0{\pm}7.0$   & $4.7{\pm}8.2$   \\
  & back    & N/A   & & 0.0   & & $57.7{\pm}29.0$ & $60.4{\pm}27.6$ & $62.1{\pm}27.6$
                                & & $83.7{\pm}23.7$ & $82.9{\pm}23.4$ & $80.3{\pm}17.3$ \\
  & none    & 94.4  & & N/A   & & N/A & N/A & N/A & & N/A & N/A & N/A \\
  & both    & 100.0 & & N/A   & & N/A & N/A & N/A & & N/A & N/A & N/A \\
\cmidrule{2-13}
  & \textit{Overall} & \textit{81.8} & & \textit{2.3}
    & & \textit{$54.5{\pm}10.9$} & \textit{$55.8{\pm}10.2$} & \textit{$55.5{\pm}10.3$}
    & & \textit{$70.1{\pm}10.1$} & \textit{$69.9{\pm}9.9$} & \textit{$69.2{\pm}7.4$} \\
\midrule
\multirow{4}{*}{\rotatebox[origin=c]{90}{Count}}
  & 0 vehicles & 94.4 & & N/A  & & N/A & N/A & N/A & & N/A & N/A & N/A \\
  & 1 vehicle  & 95.5 & & 7.8  & & $94.8{\pm}4.0$  & $95.7{\pm}3.8$  & $93.7{\pm}8.2$
                               & & $94.8{\pm}3.4$  & $95.4{\pm}2.7$  & $93.6{\pm}9.0$  \\
  & 2 vehicles & 93.8 & & 9.3  & & $49.9{\pm}24.0$ & $47.3{\pm}26.0$ & $38.4{\pm}33.5$
                               & & $59.0{\pm}16.6$ & $58.8{\pm}17.9$ & $48.5{\pm}27.0$ \\
\cmidrule{2-13}
  & \textit{Overall} & \textit{94.5} & & \textit{8.2}
    & & \textit{$83.2{\pm}4.5$}  & \textit{$83.2{\pm}5.1$}  & \textit{$79.4{\pm}4.7$}
    & & \textit{$85.7{\pm}3.1$}  & \textit{$86.2{\pm}3.7$}  & \textit{$82.1{\pm}6.2$}  \\
\midrule
\multirow{5}{*}{\rotatebox[origin=c]{90}{Type}}
  & car     & 85.7 & & 9.6  & & $52.5{\pm}31.3$ & $52.7{\pm}31.9$ & $53.3{\pm}33.6$
                            & & $57.3{\pm}20.2$ & $57.8{\pm}21.7$ & $53.3{\pm}25.5$ \\
  & van     & 97.0 & & 0.4  & & $54.0{\pm}32.7$ & $55.0{\pm}33.6$ & $51.1{\pm}36.2$
                            & & $53.7{\pm}17.1$ & $53.2{\pm}18.5$ & $55.0{\pm}21.3$ \\
  & car+van & 94.2 & & 8.4  & & $51.5{\pm}21.7$ & $48.9{\pm}24.5$ & $38.6{\pm}32.3$
                            & & $62.2{\pm}16.0$ & $61.6{\pm}17.8$ & $50.0{\pm}26.3$ \\
  & none    & 94.7 & & N/A  & & N/A & N/A & N/A & & N/A & N/A & N/A \\
\cmidrule{2-13}
  & \textit{Overall} & \textit{93.7} & & \textit{6.5}
    & & \textit{$52.1{\pm}7.8$}  & \textit{$51.8{\pm}7.9$}  & \textit{$48.2{\pm}9.7$}
    & & \textit{$56.8{\pm}3.6$}  & \textit{$56.7{\pm}3.9$}  & \textit{$52.2{\pm}4.3$}  \\
\bottomrule
\end{tabular}
\end{table*}

\subsection{Qualitative Analysis}

\subsubsection{Detection horizon outside the Field-of-View (FoV)}
For each of the $38$ validation scenarios where vehicles are entering the FoV approaching it from the left/right, the \textit{detection horizon} is defined as the longest run of consecutive $1$s frames classified correctly, ending at the frame immediately preceding the vehicle's entry into the camera's FoV.  A single frame classified differently terminates the run, so the metric is conservative. The model achieves a mean detection horizon of $4.6$ s (median $5.0$ s), with $77\%$ of scenarios providing at least $1$s of advance prediction and $62\%$ at least $3$s. Those results further highlight the benefits in the use of sound to \textit{anticipate} the presence of a traffic agent before it becomes visible, allowing for longer response times and increased safety.   

\begin{figure}[t]
        \centering
        \includegraphics[width=\linewidth]{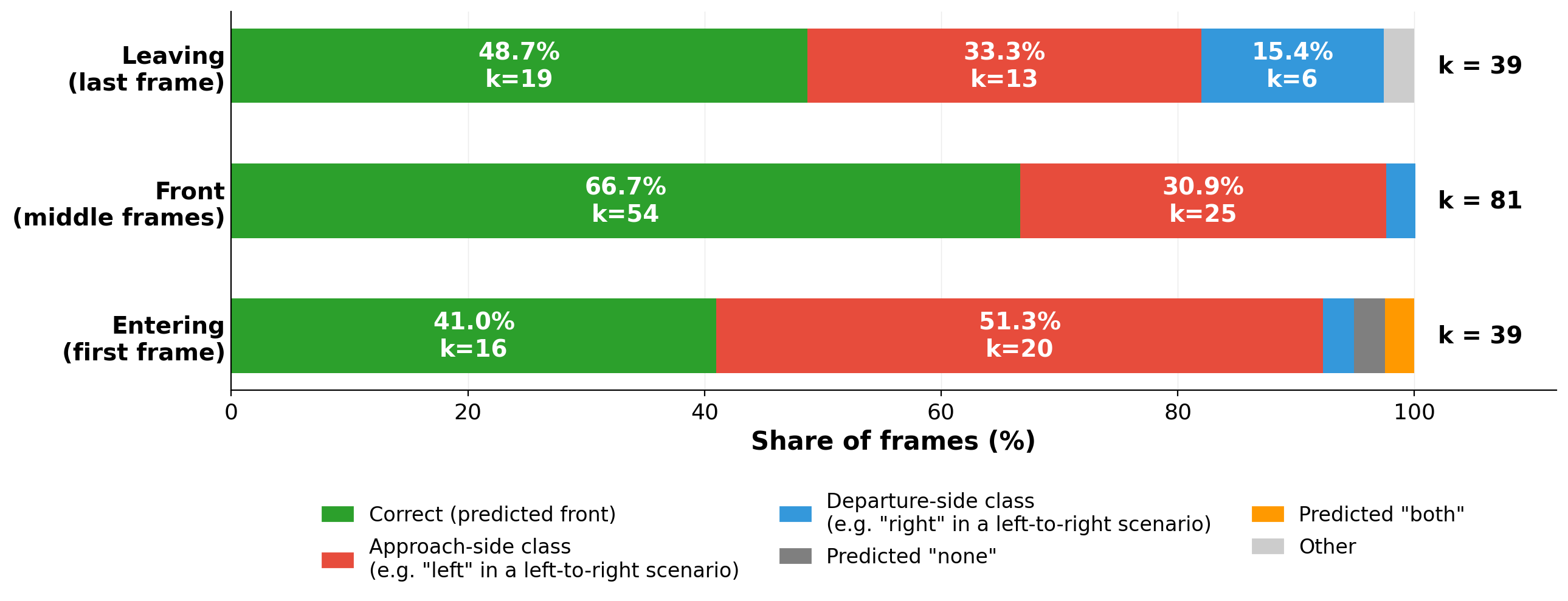}
    \caption{Predictions on the \textit{front} class in the \textit{Direction task}, broken down by frame position within the \textit{front visibility cone} (cf., Section \ref{subsec:front_error_analysis}); each row reports the share of predictions in each subclass. \textit{k} the number of frames.}
    \label{fig:front_pathology}
\end{figure}

\subsubsection{Front-class error pathology}
\label{subsec:front_error_analysis}

In \textit{Experiment 1}  the \textit{front} class yield $60.0\%$ recall. In an attempt to investigate the reasons making it the hardest to recognise, we analyse \textit{when} most errors take place. The intuition is that the \textit{Direction} classes, identified by the cone described in Figure \ref{fig:carsetting} are rather \textit{arbitrary} and not acoustically grounded, thus mistakes in the transitions are \textit{physiological}. We divide the class into three subclasses: \textit{entering, front, leaving}. The \textit{entering} subclass contains all samples corresponding to the first $1$s frame where the vehicle just becomes visible when \textit{entering} the ``front visible cone'' from left or right. The \textit{leaving} subclass contains all samples corresponding to the last $1$s frame where the vehicle is still visible before \textit{leaving} from the ``front visible cone'' heading left or right. The \textit{front} subclass contains all the remaining samples from the original \textit{front} class. Fig. \ref{fig:front_pathology} reports the results of the analysis. As we can see, most mistakes happen when ``transitioning'' from the \textit{left}/\textit{right} classes to the \textit{front} and viceversa, supporting the initial intuition. 

\subsubsection{Dynamics and Types Representations}
\label{subsec:structure}

We analyse the SSL latent abilities to capture jointly the dynamics and the type of the sound sources, and ask what the representation actually contains without supervision. Figure \ref{fig:umap_three} shows a UMAP \cite{McInnes2018UMAP} projection of the learned latents. Three structural facts are visible. First, silence consistently lives outside the active manifold: the \textit{none} class of direction, the \textit{0} class of count, and the \textit{none} class of type occupy the same peripheral islands, indicating a stable \textit{no vehicle} state separated from any vehicle activity. Second, the dominant axis within the active manifold separates one-vehicle from two-vehicle scenarios. Figure \ref{fig:umap_three}-b shows a clean spatial gradient with two-vehicle frames (\textit{2}) on the left and one-vehicle frames (\textit{1}) on the right. Figure \ref{fig:umap_three}-c inherits this structure since two-vehicle scenarios are always car+van pairs by recording protocol. Third, within the one-vehicle subset of \ref{fig:umap_three}-c, \textit{car} is less clustered than \textit{van}, demonstrating a weaker representation of this class. Figure \ref{fig:umap_three}-a shows that direction-of-arrival classes form recognisable regions: \textit{left} (blue) and \textit{right} (green) occupy opposite sides of the active manifold, \textit{both} (orange) clusters tightly on the left edge where the two-vehicle region lies, and \textit{front} (yellow) intermixes with the others, partially in coherence with the two opposite sides, but also highlighting the encoder's limits to capture the spatial structure of that class.

\begin{figure}[t]
    \centering
    \begin{minipage}[t]{0.34\columnwidth}
        \centering
        \includegraphics[width=\linewidth]{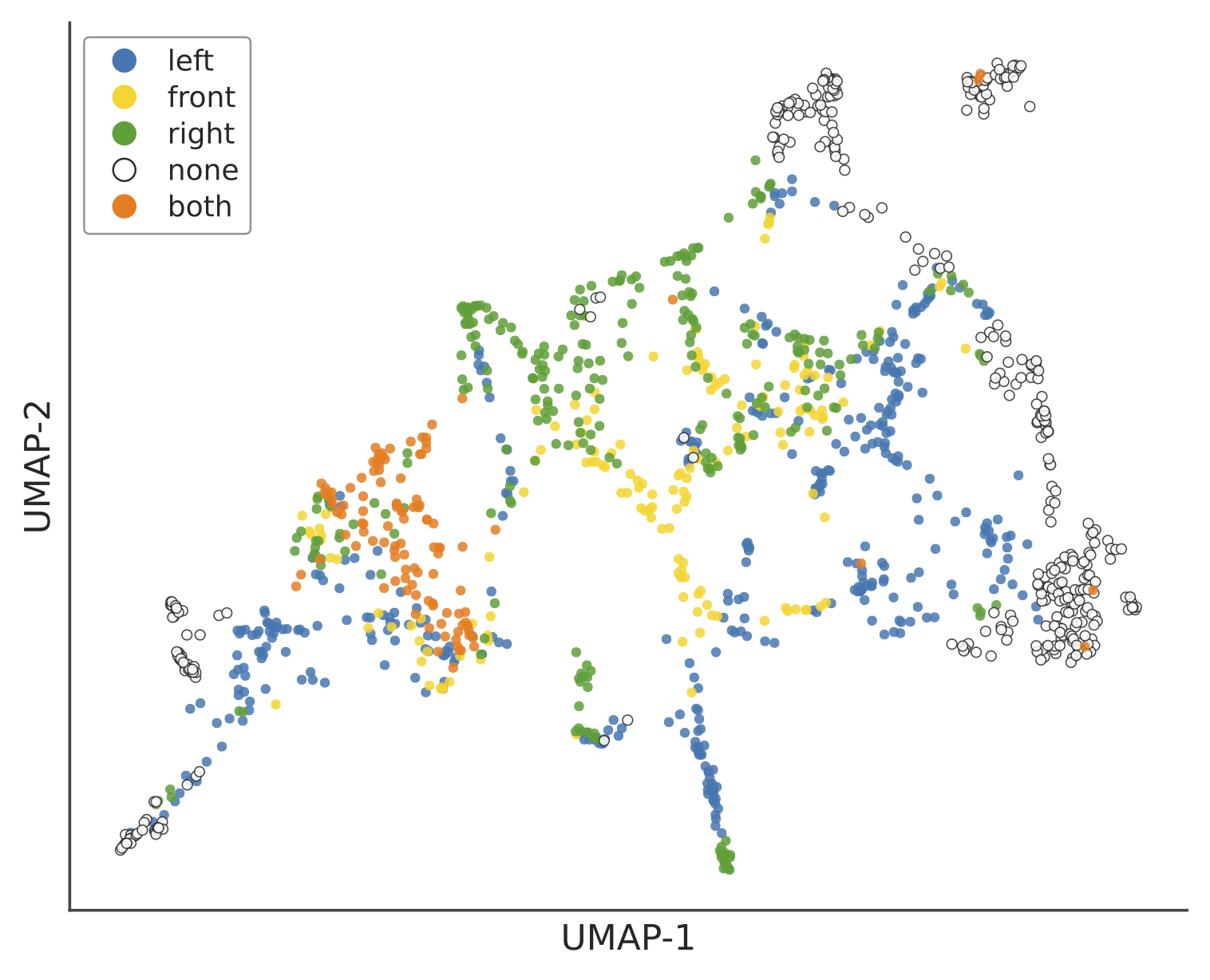}\\
        {\small (a)}
    \end{minipage}\hfill
    \begin{minipage}[t]{0.32\columnwidth}
        \centering
        \includegraphics[width=\linewidth]{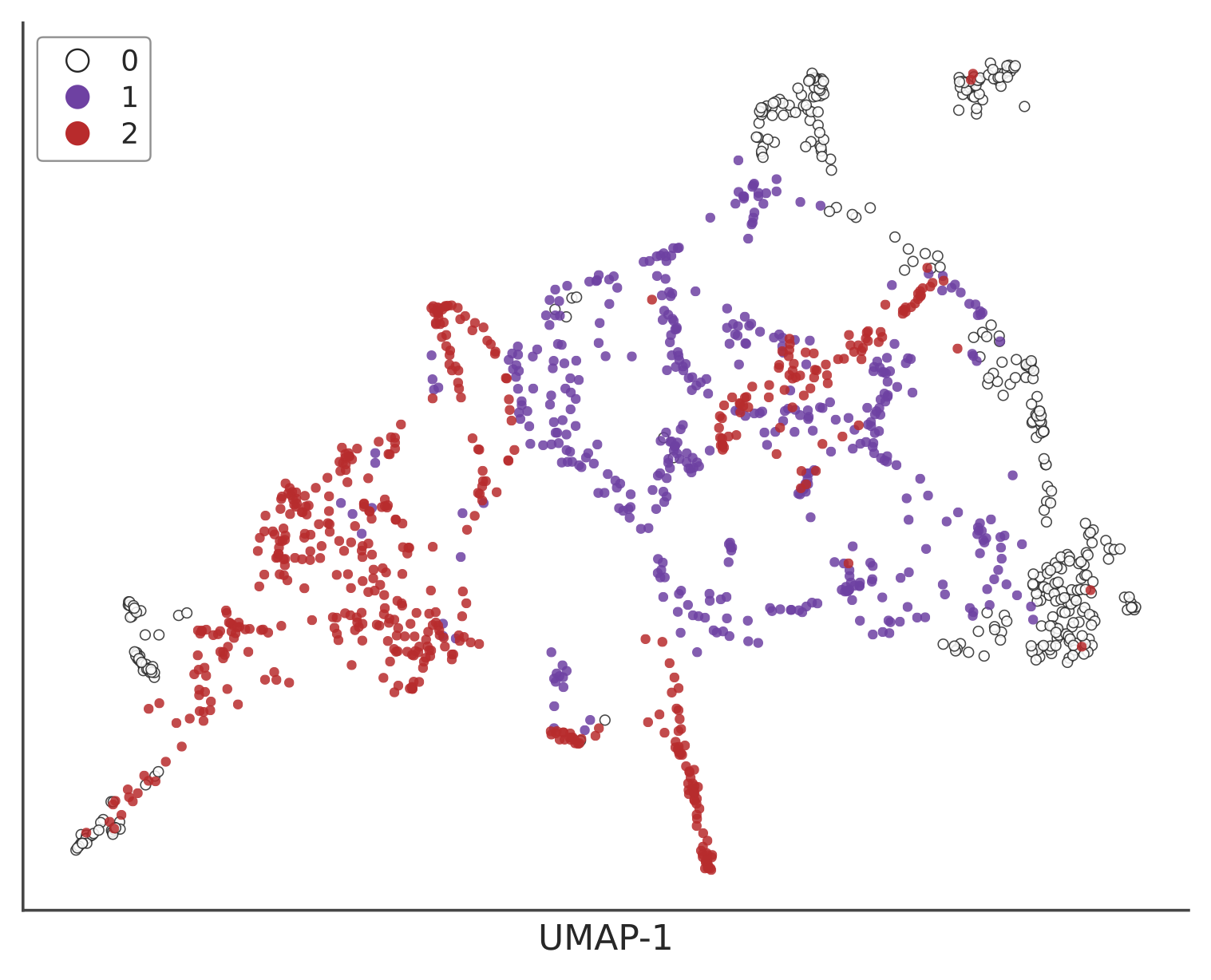}\\
        {\small (b)}
    \end{minipage}\hfill
    \begin{minipage}[t]{0.32\columnwidth}
        \centering
        \includegraphics[width=\linewidth]{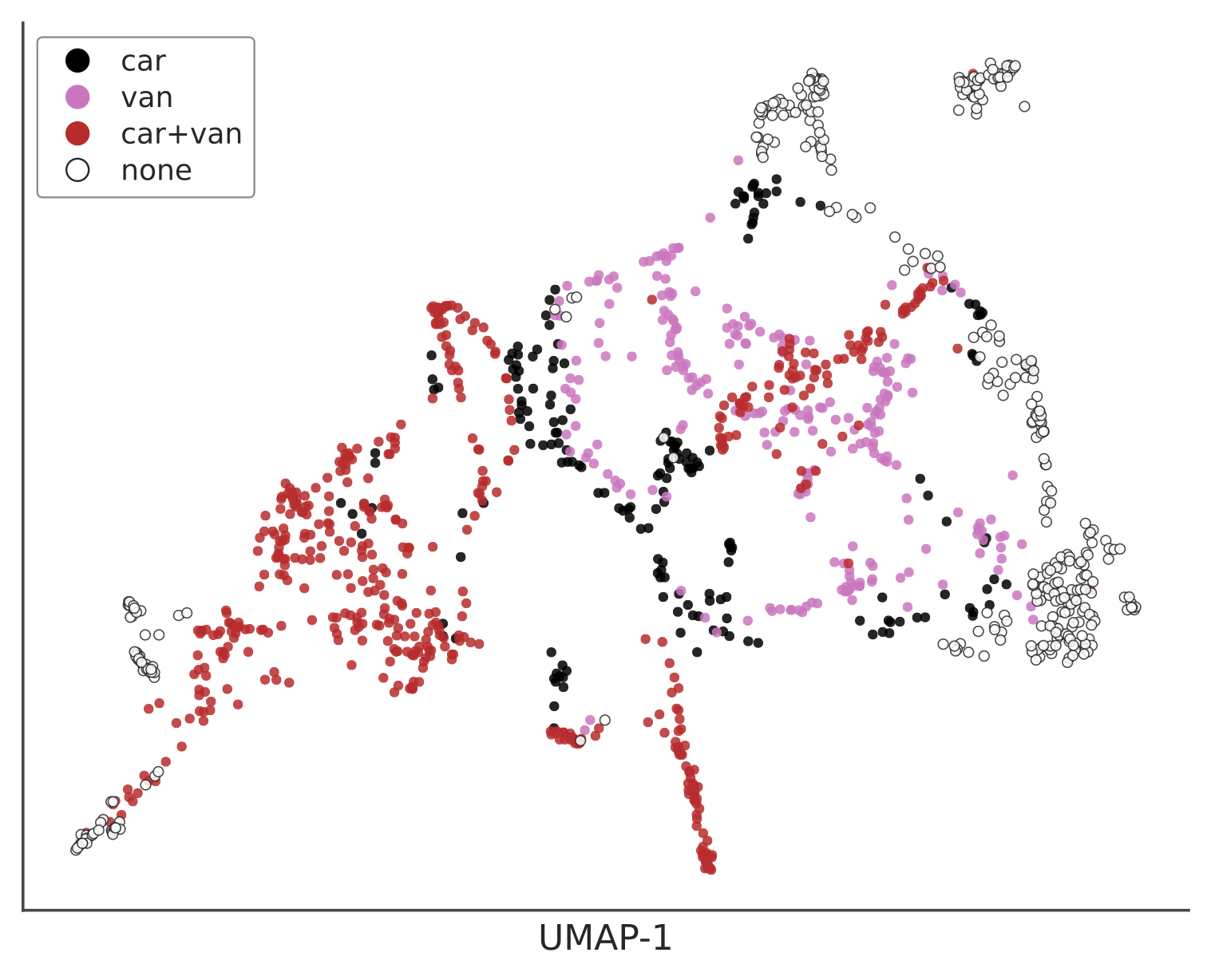}\\
        {\small (c)}
    \end{minipage}
    \caption{\small UMAP projection of the frozen-encoder validation latents (same projection coordinates across all panels), coloured by the three downstream classification labels. (a) Direction (b) Count (c) Type.}
    \label{fig:umap_three}
\end{figure}


\section{CONCLUSION}
We introduced a multi-task acoustic perception framework that estimates the direction, the number, and the type of \textit{occluded} vehicles in the scene. The framework works directly on the raw audio signals and provides robust predictions every second. Experimental evaluation showed that our method outperforms the state of the art, and further stressed the crucial role that auditory perception can play. Future work might proceed along two main directions: the extension to scenarios with more than 2 vehicles present, where localisation  is refined to a regression task (\textit{i.e.}, from classes to continuous angles), and the integration into a multimodal system.

\section{ACKNOWLEDGEMENT}
This work was supported by the Fraunhofer IDMT institute for the materials to do the data collection part. Thanks to the \textit{hearing car} team and employees from the institute, we were able to collect data in a controlled and safe environment for the experiments.

\bibliographystyle{IEEEtran}
\bibliography{references}

\end{document}